%% file: main.tex
\documentclass[11pt]{article}
\usepackage[margin=1in]{geometry}
\usepackage[T1]{fontenc}
\usepackage{microtype}
\usepackage{amsmath,amssymb,amsthm,mathtools}
\usepackage[sc,osf]{mathpazo}
\AddToHook{env/table/begin}{\fontfamily{pplx}\selectfont}
\usepackage{booktabs,graphicx,array}
\usepackage{placeins}
\usepackage[round,authoryear]{natbib}
\usepackage[hidelinks]{hyperref}
\newcommand{\E}{\mathbb E}
\newcommand{\Prob}{\mathbb P}

\newcommand{\Cov}{\operatorname{Cov}}
\DeclareMathOperator*{\argmin}{arg\,min}
\DeclareMathOperator{\diag}{diag}
\newtheorem{proposition}{Proposition}
\ifdefined\pdfinfoomitdate
  \pdftrailerid{}
\fi
\hypersetup{pdftitle={Count Constraints and Conditional Benchmarks for Roster-Based Network Reconstruction},pdfauthor={Yen-hsuan Tseng}}
\title{Count Constraints and Conditional Benchmarks for\\
Roster-Based Network Reconstruction}
\author{Yen-hsuan Tseng}
\date{}
\begin{document}
\maketitle
\begin{abstract}
Complete roster-based aggregated relational data give each person's number of contacts in each known group. We compare reconstruction methods on pairs whose edge states these counts leave unresolved. In a sparse circular latent-position simulation with 40 nodes and eight groups, the counts fix 35.6\% of realized edges and 91.7\% of nonedges on average. Ranking forced edges first and forced nonedges last, with all unresolved pairs tied, yields a mean area under the receiver operating characteristic curve (AUC) of 0.973 across all pairs. The uniform reference uses only the roster and counts, giving equal weight to all compatible graphs. On unresolved pairs its mean AUC exceeds that of both the direct circular and nuclear-norm scores. In a post hoc analysis of the same graphs, weighting by the fitted circular odds reverses this comparison. With the generating form and interaction strength supplied, the circular plug-in score attains mean AUC 0.696 versus 0.681 for uniform in the sparse design. The gain over uniform is 0.016--0.019 across three coupled circular densities and 0.007 under a stochastic blockmodel, where the circular form is misspecified. Nuclear-norm plug-in scores yield lower mean AUC than uniform weighting. For independent Bernoulli edges with fixed interior probabilities, weighting compatible graphs by the true generating odds (the conditional oracle) gives the same edge marginals as uniform weighting for every feasible count array exactly when log odds are additive in node-by-group effects. We recommend reporting fixed edge states, ranking on unresolved pairs, and generating-probability error separately.
\end{abstract}


\section{Introduction}\label{sec:intro}
Complete roster-based aggregated relational data (ARD) can reveal who is connected to whom. Each respondent reports how many contacts they have in each known group. A zero count rules out every link from that respondent to that group. Positive counts can also determine individual ties. The first two rows in Figure~\ref{fig:positive-margins} require six ties in total. At most two can fit in the last two columns, so all four entries in the upper-left block must be one. We call the pairs for which either state remains possible the unresolved set $\mathcal V$. How well does a fitted model rank ties within that set?
\begin{figure}[htbp]
\centering
\[
\left[\begin{array}{cc|cc}
1&1&\ast&\ast\\
1&1&\ast&\ast\\\hline
\ast&\ast&0&0\\
\ast&\ast&0&0
\end{array}\right]
\qquad
\begin{array}{l}
\text{Row sums: }(3,3,1,1)\\[3pt]
\text{Column sums: }(3,3,1,1)
\end{array}
\]
\caption{Complete counts can fix edge states even when no counts are zero. This is a between-group block for two distinct groups of four nodes. All margins are positive and none is full, yet the upper-left block is forced to one and the lower-right block to zero. Each starred $2\times2$ block must be a permutation matrix, giving four compatible matrices in total. Stars denote states that can vary, subject to those joint constraints.}
\label{fig:positive-margins}
\end{figure}

Latent-position models express tie probabilities through distances between nodes \citep{hoff2002,mccormick2015}. Penalized matrix estimation offers a second approach to reconstruction from group counts \citep{alidaee2020}. In a comparison on four simulated networks with the same counts available to both methods, the nuclear-norm fit ranked realized edges more accurately and matched the counts more closely, while the circular latent-position fit estimated the generating probabilities more accurately \citep{tseng2025comparison}. Here we compare both fits with a reference that gives every graph compatible with the counts equal weight. We score unresolved pairs directly from each fitted matrix and then examine what its odds contribute when used to weight those same graphs.

\begin{samepage}
Our forty exploratory graphs include the four from the earlier study. We then evaluated prespecified comparisons on a fresh sample of forty circular replicate blocks, each covering all three densities, and on a separate sample of forty stochastic blockmodel graphs. After examining the direct scores, we used the saved fits from all five settings for the plug-in comparison.
\begin{enumerate}
\item \textit{States fixed by the counts.} In the fresh sparse eight-group design, the counts fix 35.6\% of realized edges and 91.7\% of nonedges on average. Assigning scores of $1$, $1/2$ and $0$ to forced edges, unresolved pairs and forced nonedges, respectively, gives a mean full-pair AUC of 0.973. On the unresolved pairs alone, the same score has AUC 0.5 on every graph.
\item \textit{When generating odds matter.} For a fixed independent-edge Bernoulli law with interior probabilities, Proposition~\ref{prop:conditional-uniform} characterizes agreement between the oracle and uniform edge marginals on every feasible count array. The condition is the node-by-group additive log-odds form of the fixed-label $k$-beta model of \citet{csiszar2012testing}.
\item \textit{Direct and plug-in scores.} Uniform weighting gives higher mean unresolved-pair AUC than the direct circular and nuclear scores in the fresh circular and blockmodel studies. Weighting compatible graphs by the saved circular fits raises their mean AUC above uniform. Nuclear plug-in scores remain below uniform at all three penalties.
\end{enumerate}
\end{samepage}

Inference about the generating model uses ARD in several ways. Combining survey counts with census data permits estimation of a network formation model \citep{breza2020}, with consistency results for parameters and network statistics under stated conditions \citep{breza2023}. Spectral tests can assess model fit from complete and partial network data, including ARD \citep{lubold2025}. Identifying a generating probability matrix from the population count law still leaves the problem of estimating it from one network. In a weak-signal logistic model with a fixed number of equal-sized observation groups, the worst-case risk for estimating the matrix stays bounded away from zero as group size grows \citep[Corollary~2.4]{tseng2026impossibility}. Our simulations evaluate finite-sample error against the known generating probabilities as well as ranking accuracy for the realized graph.

What a reconstruction can recover depends on which observations are available. With missing or noisy adjacency entries, \citet{guimera2009} distinguish model-based link reliabilities from heuristic scores, \citet{peixoto2018} infers networks conditional on noisy observations, and \citet{gaucherklopp2021} study probability-matrix estimation from partially observed adjacency matrices. \citet{neal2024} infers ties from records of who belongs to which groups. Shared memberships provide a score for a possible tie even when no contact counts are collected. \citet{januar2026} show how preferential edge sampling can produce apparent structure. In roster ARD, each person's reported number of ties to a group fixes a sum of edge indicators. \citet{davisonward2026} study recovery of latent communities from aggregate counts under degree-corrected stochastic blockmodels. Here we hold the observed counts fixed, identify node pairs whose edge state is the same in every compatible graph, and compare rankings of the remaining pairs.

Conditional distributions under fixed margins provide our benchmarks. Independent Bernoulli entries conditioned on row and column sums weight each compatible matrix by a product of odds \citep{harrison2013weighted}. Exact counting and uniform sampling for fixed-margin matrices are developed by \citet{miller2013exact}. We enumerate the small blocks of the eight-group design to evaluate rankings on the same unresolved pairs, first with direct scores and then with the fitted odds as weights.

\section{Observation model and evaluation targets}\label{sec:observation}
Let \(A=(a_{ij})\) be a symmetric binary adjacency matrix with zero diagonal. A known matrix \(B\in\{0,1\}^{n\times K}\) records mutually exclusive groups \(G_1,\ldots,G_K\) that cover the nodes. Every node reports
\begin{equation}
 y_{ik}=\sum_{j\ne i}B_{jk}a_{ij},\qquad Y=AB.
 \label{eq:ard}
\end{equation}
The roster supplies the group memberships in $B$, while the latent positions used in the circular model remain unobserved. Observing the full array $Y$ also requires a count report from each roster member. For comparison, \citet{breza2020} sampled households for their village surveys and combined their reports with census information.

Conditional on a fixed generating matrix \(P^\star\), the unordered edge indicators are independent Bernoulli variables with probabilities \(p^\star_{ij}\in(0,1)\). A fitted symmetric probability matrix with zero diagonal predicts count means
\begin{equation}
 \mu_{ik}(\theta)=\sum_{j\ne i}B_{jk}p_{ij}(\theta),\qquad M(\theta)=P(\theta)B.
 \label{eq:mean}
\end{equation}
Given $P^\star$ and $B$, the expected count matrix is \(P^\star B\). Two respondents' counts covary when both include their shared edge, as follows from edge independence.
\begin{equation}
 \Cov(y_{ik},y_{j\ell}\mid P^\star,B)
 =B_{jk}B_{i\ell}p^\star_{ij}(1-p^\star_{ij}),\qquad i\ne j.
 \label{eq:dependence}
\end{equation}

We evaluate edge rankings and Brier loss against the realized adjacency matrix $A$, probability root mean squared error (RMSE) against $P^\star$, and count RMSE against the observed $Y$. Count RMSE measures in-sample fit. The ranking target is the particular graph whose edges contributed to $Y$. For a new graph drawn independently from the same fixed $P^\star$, observing $Y$ leaves the probability of edge $ij$ at $p^\star_{ij}$.

\section{Fitted representations and count scores}\label{sec:methods}
\subsection{Circular working posterior}
The circular representation is
\begin{equation}
 p_{ij}(\alpha,t)=\sigma\{\alpha+\zeta\cos(t_i-t_j)\},\qquad
 \sigma(x)=(1+\exp(-x))^{-1},\quad i\ne j,
 \label{eq:circle}
\end{equation}
with diagonal zero. In the circular experiments we fix $\zeta$ at its generating value and estimate the intercept and latent positions. We use the same form and strength in the blockmodel experiment, where the circular form is misspecified. In both designs the fit uses the working log-likelihood
\begin{equation}
 \ell_W(\alpha,t)=\sum_{i,k}\{y_{ik}\log\mu_{ik}(\alpha,t)-\mu_{ik}(\alpha,t)-\log(y_{ik}!)\}.
 \label{eq:working}
\end{equation}
Conditional on $P^\star$, each count has a Poisson-binomial distribution, and counts that share an edge can be dependent by \eqref{eq:dependence}. These features make the product-Poisson likelihood a working approximation even in the circular experiments. Combining it with a prior gives \(\pi_W(\alpha,t\mid Y,B)\propto\exp\{\ell_W(\alpha,t)\}\pi(\alpha,t)\).

The fitted matrix averages edge probabilities over retained draws. These probabilities are invariant to rotations and reflections of the angles, so their posterior mean can be computed without aligning the draws.

\subsection{Nuclear-norm probability matrix}
The second representation is defined by the constrained problem
\begin{equation}
 \widehat P\in\argmin_{P\in\mathcal C}
 \left\{\tfrac12\|PB-Y\|_F^2+\lambda\|P\|_*\right\},\qquad
 \mathcal C=\{P=P^\top,\ \diag P=0,\ 0\le P\le1\}.
 \label{eq:nuclear}
\end{equation}
This adapts the probability-matrix formulation of \citet[eq.~(5)]{alidaee2020} to symmetric matrices with zero diagonal. We specified \(\lambda=1\) as the primary penalty and report all three values on the fixed grid \(\{0.25,1,4\}\). The same raw penalty has a different strength relative to the fit term under the four- and eight-group partitions ($K=4,8$). The random-trait and penalty conditions of the cited recovery guarantees are not imposed here.

The fit term measures the discrepancy between predicted count means and the observed counts. The penalty trades that discrepancy against the nuclear norm, so the fitted matrix can have $\widehat P B\ne Y$ even when the optimizer converges.

\subsection{Density, count product and zero exclusion}
Since groups partition the roster and all nodes report, the observed counts identify the realized density,
\begin{equation}
 \widehat p=\frac{\sum_{i,k}y_{ik}}{n(n-1)}.
 \label{eq:density}
\end{equation}
The density baseline assigns this value to all off-diagonal entries. A parameter-free ranking score is
\begin{equation}
 S_{ij}=Y_{i,g(j)}Y_{j,g(i)},\qquad
 Z_{ij}=\mathbf1\{S_{ij}>0\},\qquad
 \mathcal U=\{(i,j):i<j,\ Z_{ij}=1\},
 \label{eq:product}
\end{equation}
where \(g(i)\) is the group of node \(i\). The product ranks pairs by the two observable counts. Either count being zero implies \(a_{ij}=0\), so every realized edge lies in \(\mathcal U\).

Let \(q_0\) be the proportion of nonedges excluded by this zero rule. The binary score \(Z\) has
\begin{equation}
 \operatorname{AUC}(Z,A)=\tfrac12+\tfrac12q_0.
 \label{eq:zero}
\end{equation}
For any nonnegative score \(T\), define its zero-adjusted version \(T^Z=T Z\). If \(r_0\) is the proportion of realized edges receiving score zero, then, whenever \(\mathcal U\) contains both classes,
\begin{equation}
 \operatorname{AUC}(T^Z,A)
 =q_0(1-\tfrac12r_0)+(1-q_0)\operatorname{AUC}_{\mathcal U}(T,A).
 \label{eq:zero-decomposition}
\end{equation}
A positive edge score beats every excluded nonedge, and an edge scored zero ties with those nonedges. Within \(\mathcal U\), zero adjustment preserves all pairwise comparisons. We give ties half credit and evaluate the identity on each graph before averaging.

\section{Complete constraints and conditional references}\label{sec:conditional}
\subsection{Feasible graphs and resolved pairs}
We call the set of graphs compatible with a given count array its \emph{fibre},
\begin{equation}
 \mathcal F(Y,B)=
 \{a:a=a^\top,\ a_{ii}=0,\ a_{ij}\in\{0,1\},\ aB=Y\}.
 \label{eq:fibre}
\end{equation}
Define \(\mathcal F_0\) and \(\mathcal F_1\) as the unordered pairs whose entries equal zero and one, respectively, in every member of \(\mathcal F\). The remaining pairs form
\begin{equation}
 \mathcal V=\{(i,j):i<j\}\setminus(\mathcal F_0\cup\mathcal F_1).
 \label{eq:unresolved}
\end{equation}
Zero-excluded pairs belong to \(\mathcal F_0\), hence \(\mathcal V\subseteq\mathcal U\). The inclusion can be strict: Figure~\ref{fig:positive-margins} has positive margins but fixes both absent and present edges through the complete constraints.

For \(g<h\), the observed counts give every row and column sum of the between-group block \(A_{G_g,G_h}\). For \(g=h\), they give the degree sequence of the within-group simple graph. These constraints involve disjoint unordered edge sets, so
\begin{equation}
 \mathcal F(Y,B)=\prod_{g\le h}\mathcal F_{gh}.
 \label{eq:fibre-product}
\end{equation}
For the balanced eight-group design, each group contains five nodes. For a between-group block with row sums $y_{ih}$ for observation group $h$, there are at most \(\prod_i\binom5{y_{ih}}\le10^5\) candidate row patterns before column-sum filtering; a within-group block has \(2^{10}\) possible edge configurations. Exact enumeration is therefore practical for each block in this design.

\subsection{Uniform reference and conditional oracle}
Our two conditional benchmarks assign an edge probability to each pair,
\begin{equation}
 Q^{\mathrm{unif}}_{ij}=\E_{\mathrm{Unif}(\mathcal F)}[A_{ij}],
 \qquad
 Q^{\mathrm{or}}_{ij}=\Prob(A_{ij}=1\mid Y,B,P^\star).
 \label{eq:references}
\end{equation}
Equal weighting specifies a reference law on the graphs compatible with $(Y,B)$. Its edge marginals can be computed from these observations alone. The oracle additionally uses the generating $P^\star$ to weight the graphs.

With \(w_{ij}=p^\star_{ij}/(1-p^\star_{ij})\), independence before conditioning gives the established fixed-margin weighted law \citep{harrison2013weighted},
\begin{equation}
 \Prob(A=a\mid Y,B,P^\star)
 \propto \mathbf1\{a\in\mathcal F\}\prod_{i<j}w_{ij}^{a_{ij}}.
 \label{eq:weighted}
\end{equation}
Together with \eqref{eq:fibre-product}, this factors the conditional distribution and its normalizing constant across unordered group-pair blocks. We enumerate each small support directly, count edge occurrences as integers for \(Q^{\mathrm{unif}}\), and use log-shifted floating-point weighted sums for \(Q^{\mathrm{or}}\). The enumeration uses the conditioning structure of fixed-margin methods \citep{harrison2013weighted,miller2013exact}.

Both references satisfy \(QB=Y\). The fitted methods use the counts to estimate generating probabilities, but their reported matrices have not been conditioned on $\mathcal F(Y,B)$. Comparing these direct scores with the references therefore combines differences in fitted odds with differences in the use of the exact constraints. Strictly interior generating probabilities give both references the same support and forced states. We identify forced states using exact integer support counts. The factorization above conditions on fixed \(P^\star\); mixing over unknown latent positions can couple the blocks. Proposition~\ref{prop:conditional-uniform} characterizes the fixed generating matrices for which the conditional law is uniform on every feasible count fibre.

Some model structure disappears after conditioning. For a between-group block, additive log odds $\log w_{ij}=r_i+c_j$ have a constant sum over the edges of every graph in the fixed-margin support \citep[sec.~2.2]{harrison2013weighted}. In the circular model, the intercept likewise contributes a constant because each block's edge count is fixed. The gap between the two references therefore concerns odds structure that remains after conditioning.

\input{conditional-uniformity.tex}

\subsection{Conditional loss and ranking}
For a probability score \(T\), let \(\operatorname{BS}_D(T,A)\) be the mean squared error against realized edges over a nonempty pair set \(D\). Conditional on \(Y,B,P^\star\), a fixed score and a pair set determined by these counts satisfy
\begin{equation}
 \E[\operatorname{BS}_D(T,A)\mid Y,B,P^\star]
 =\frac1{|D|}\sum_{e\in D}
 \left\{Q^{\mathrm{or}}_e(1-Q^{\mathrm{or}}_e)
 +(T_e-Q^{\mathrm{or}}_e)^2\right\}.
 \label{eq:brier}
\end{equation}
Thus the mean squared distance from \(Q^{\mathrm{or}}\) is conditional excess Brier risk for reconstructing \(A\). The identity holds even when edges are dependent after conditioning. Generating-probability error is assessed separately against \(P^\star\).

The oracle also orders pairs optimally for conditional expected AUC on the full set, \(\mathcal U\), or \(\mathcal V\), provided both classes are present. In each case, the counts fix the class totals. For two distinct pairs \(e,f\),
\begin{equation}
 \Prob(A_e=1,A_f=0\mid Y,B,P^\star)
 -\Prob(A_f=1,A_e=0\mid Y,B,P^\star)
 =Q^{\mathrm{or}}_e-Q^{\mathrm{or}}_f.
 \label{eq:auc-order}
\end{equation}
Ranking by the oracle maximizes each conditional expected pair contribution with a fixed denominator. On an individual realized graph, another score can have higher empirical AUC.

Let \(m,n_0\) be the full-set numbers of edges and nonedges, and \(m_{\mathcal V},n_{\mathcal V}\) their numbers in \(\mathcal V\). A score that places forced ones above all unresolved pairs and forced zeros below them has
\begin{equation}
 \operatorname{AUC}_{\mathrm{full}}
 =1-\frac{m_{\mathcal V}n_{\mathcal V}}{mn_0}
   \{1-\operatorname{AUC}_{\mathcal V}\}.
 \label{eq:full-decomposition}
\end{equation}
Only edge--nonedge comparisons within \(\mathcal V\) can lower this AUC. The two conditional references have this ordering, since their unresolved probabilities are strictly between zero and one. For any other score, placing forced zeros below $\mathcal V$ and forced ones above it, while retaining the original ordering and ties within $\mathcal V$, gives the same decomposition. The forced-state-only score instead ties all pairs within $\mathcal V$, so its AUC is $1-m_{\mathcal V}n_{\mathcal V}/(2mn_0)$.

\input{plugin-method.tex}

\section{Empirical evidence}\label{sec:empirical}
\subsection{Evaluation}\label{sec:evaluation}
\input{evaluation.tex}

\subsection{Fresh circular samples at three densities}\label{sec:fresh}
\input{density-study.tex}
\FloatBarrier

\subsection{A matched stochastic blockmodel contrast}\label{sec:sbm}
\input{sbm-study.tex}

\subsection{Exploratory context}\label{sec:numerical}
\input{exploratory-summary.tex}

\subsection{Plug-in conditional comparison}\label{sec:plugin}
\input{plugin-study.tex}
\FloatBarrier

\section{Discussion}\label{sec:discussion}
A score that ties every unresolved pair still attains mean full-pair AUC 0.973 in the fresh sparse eight-group design. It earns this score by placing forced edges first and forced nonedges last. Equation~\eqref{eq:full-decomposition} shows why the overall AUC changes so little when the ranking improves within $\mathcal V$. Only comparisons between an edge and a nonedge that both belong to $\mathcal V$ distinguish scores with the same forced-state ordering.

On the unresolved pairs in the fresh sparse setting, the circular fit gives mean AUC 0.582 when its entries score pairs directly. Using those same probabilities to weight compatible graphs raises mean AUC to 0.696, above the uniform reference at 0.681. Conditioning the nuclear fit at $\lambda=1$ also improves its mean AUC in this setting, from 0.636 to 0.668, but uniform still ranks higher. For circular, the direct-to-plug-in difference measures the change in ranking when the same fitted probabilities are conditioned on the counts. For nuclear, the plug-in calculation first clips the fitted probabilities. Clipping alone leaves the reported mean AUC on $\mathcal V$ unchanged at the primary threshold (Appendix~\ref{app:plugin-checks}). Comparing the plug-in score with uniform measures what the fitted odds contribute once both scores use all the counts.

The fresh sample reproduces the trade-off seen in the earlier matched-input comparison \citep{tseng2025comparison}. In the fresh sparse eight-group setting, circular has lower mean RMSE against $P^\star$ than nuclear at $\lambda=1$ (0.0698 versus 0.1325), while nuclear has higher mean direct AUC on the unresolved pairs of the realized graph (0.636 versus 0.582). Across the three densities, the circular fit's RMSE advantage over the density baseline grows, and nuclear at $\lambda=1$ exceeds the baseline RMSE under both partitions.

Proposition~\ref{prop:conditional-uniform} identifies which fixed generating matrices lose their weighting effect after conditioning on every feasible count array. For example, if each observation group lies inside one generating community of an SBM, the odds are constant within each block. Every compatible graph has the same edge total in that block and receives the same weight. In the simulated SBM the observation groups mix communities, so the true odds can distinguish graphs that reproduce exactly the same counts.
\paragraph{Scope.}
Our empirical evidence covers one circular family at three densities and one balanced blockmodel contrast, all at $n=40$. The circular samples use two nested balanced partitions; the blockmodel uses $K=8$. Density changes the count distribution, unresolved pair sets and relative penalty strength together. Moment matching across families fixes the population dyad mean and probability variance, while degree heterogeneity and conditional difficulty can differ. These comparisons do not isolate a causal effect of zero exclusion or model shape. The nuclear fits use a fixed penalty grid. The circular fits use $\zeta=1.5$ throughout. The blockmodel matches the probability mean and variance of the circular design at that strength and 15\% density, while its two community-dependent probabilities follow a different functional form. A correctly specified blockmodel estimator remains untested.

We computed complete conditional references for five-node groups ($K=8$). In larger between-group blocks, exact uniform sampling with prescribed row and column sums \citep{miller2013exact} offers a way to estimate uniform marginals. Importance sampling for weighted fixed-margin matrices \citep{harrison2013weighted} offers a corresponding route for the plug-in marginals. Within-group samplers must also enforce undirected symmetry, a zero diagonal and the observed degrees. Accuracy and cost at coarser partitions remain to be assessed.

The forty exploratory graphs motivated the direct-score comparisons. We then evaluated the prespecified comparisons on forty fresh replicate blocks paired across densities and a separate sample of forty blockmodel graphs. We developed the plug-in analysis after inspecting those results and evaluated it on the existing graphs. The plug-in comparison still needs replication on new graphs.

\begin{samepage}
\section{Conclusion}
When the fixed states can be identified, report how many edges and nonedges the counts determine and compare rankings on the remaining pairs. In simulations, report error against the known generating probabilities separately. To assess a fitted model, report its direct and plug-in scores alongside uniform weighting. The direct score shows how the fit ranks pairs on its own. Comparing its plug-in score with uniform shows how much the fitted weights add on the same set of compatible graphs.

\end{samepage}

\section*{Data and code availability}\label{sec:data}
All networks in this study were simulated. The replication archive contains the simulated networks, fitted matrices, conditional marginals, and code for all ten tables and both figures. It also provides code and instructions for regenerating the probability draws. The versioned archive is available at \url{https://doi.org/10.5281/zenodo.24245598}.

\section*{Use of AI tools}
OpenAI ChatGPT assisted with literature review and suggested the conditional benchmark. Anthropic Claude suggested score comparisons and conditioning the saved fits. Google Deep Review commented on an earlier draft. OpenAI Codex assisted with drafting, code, computational verification, replication materials and figures. The author takes responsibility for the scientific content.

\setlength{\bibsep}{3pt plus 1pt minus 1pt}
\bibliographystyle{plainnat}
\bibliography{verified}

\clearpage
\appendix
\numberwithin{equation}{section}
\setcounter{table}{0}
\renewcommand{\thetable}{A\arabic{table}}
\renewcommand{\theHtable}{appendix.\Alph{section}.\arabic{table}}
\section{Exploratory sample: protocol and detailed results}\label{app:exploratory}
\input{experiment.tex}
\clearpage
\input{numerical-details.tex}

\end{document}

%% file: conditional-uniformity.tex
\subsection{Characterization of conditional uniformity}\label{sec:uniformity}
The $k$-beta model of \citet[eq.~(5)]{csiszar2012testing} combines a $\beta$-model within each group with a Rasch model between each pair of groups \citep[Sections~2--3]{bolla2015logistic}. For a single group, \citet[Theorem~2.3]{csiszar2012degree} proved that uniformity given every feasible degree sequence characterizes the $\beta$-model. For complete roster counts, we characterize the corresponding multigroup law by equality of conditional edge marginals. The proof uses four-node matchings within groups and two-edge exchanges between groups. With fixed, strictly interior Bernoulli probabilities and independent edges, the oracle and uniform reference agree on every edge marginal for every feasible count array exactly when the odds follow the $k$-beta model with the observation groups as labels.

\begin{proposition}[Characterization of conditional uniformity]\label{prop:conditional-uniform}
Fix a known partition $B$, and let $g(i)$ denote the group containing node $i$. Conditional on one fixed $P^\star$, suppose the unordered edges of a simple undirected graph are independent Bernoulli variables with $0<p^\star_{ij}<1$ for $i<j$. Under the complete observation $Y=AB$, the following are equivalent:
\begin{enumerate}
\item[(i)] $Q^{\mathrm{or}}(y)=Q^{\mathrm{unif}}(y)$ for every $y$ with $\mathcal F(y,B)\ne\varnothing$.
\item[(ii)] For every such $y$, the conditional graph law is uniform:
\begin{equation}
 \Prob(A=a\mid Y=y,B,P^\star)
 =\frac{\mathbf1\{a\in\mathcal F(y,B)\}}{|\mathcal F(y,B)|}.
 \label{eq:conditional-uniform}
\end{equation}
\item[(iii)] There is a matrix $\gamma\in\mathbb R^{n\times K}$ such that
\begin{equation}
 \log\frac{p^\star_{ij}}{1-p^\star_{ij}}
 =\gamma_{i,g(j)}+\gamma_{j,g(i)},\qquad i<j.
 \label{eq:block-additive-odds}
\end{equation}
\end{enumerate}
\end{proposition}

\begin{proof}
Write $\eta_{ij}=\log\{p^\star_{ij}/(1-p^\star_{ij})\}$. For a fixed feasible $y$, let $q_y$ be the conditional graph law and $u_y$ the uniform law on $\mathcal F(y,B)$. By \eqref{eq:weighted},
\[
 q_y(a)=Z_y(\eta)^{-1}\exp\!\left(\sum_{i<j}\eta_{ij}a_{ij}\right),
 \qquad a\in\mathcal F(y,B),
\]
where $Z_y(\eta)$ is the normalizing sum. Both laws are strictly positive on this finite fibre. Their Kullback--Leibler divergences satisfy
\begin{align}
 D(q_y\Vert u_y)+D(u_y\Vert q_y)
 &=\sum_a\{q_y(a)-u_y(a)\}\log\frac{q_y(a)}{u_y(a)}\notag\\
 &=\sum_{i<j}\eta_{ij}
   \{Q^{\mathrm{or}}_{ij}(y)-Q^{\mathrm{unif}}_{ij}(y)\}.
 \label{eq:symmetric-kl}
\end{align}
The normalizing terms cancel. Under (i), the right side is zero, so nonnegativity of the divergences gives $q_y=u_y$. This proves (i)$\Rightarrow$(ii); the reverse implication follows by taking edge expectations.

To prove (ii)$\Rightarrow$(iii), fix distinct groups $G_g,G_h$ with $g<h$. For distinct $i,i'\in G_g$ and $j,j'\in G_h$, the counts generated by the graph containing only edges $ij$ and $i'j'$ admit exactly two graphs, the other having only $ij'$ and $i'j$. Uniformity gives
\begin{equation}
 \eta_{ij}+\eta_{i'j'}=\eta_{ij'}+\eta_{i'j}.
 \label{eq:odds-rectangle}
\end{equation}
Fix anchors $i_0\in G_g$ and $j_0\in G_h$, and set
\[
 r_i^{(gh)}=\eta_{ij_0},\qquad
 c_j^{(gh)}=\eta_{i_0j}-\eta_{i_0j_0}.
\]
Then $\eta_{ij}=r_i^{(gh)}+c_j^{(gh)}$: if $i=i_0$ or $j=j_0$, this follows directly from the definitions; otherwise, it follows from \eqref{eq:odds-rectangle}. If either group is a singleton, every pair falls into an anchor case, so the same representation holds without a rectangle restriction.

For the within-group construction, fix $G_g$ and relabel its $m$ vertices locally as $1,\ldots,m$. If $m=1$, take $v_1^{(g)}=0$; if $m=2$, take $v_1^{(g)}=v_2^{(g)}=\eta_{12}/2$. For $m\ge3$, define
\[
 v_1^{(g)}=\tfrac12(\eta_{12}+\eta_{13}-\eta_{23}),\qquad
 v_i^{(g)}=\eta_{1i}-v_1^{(g)}\quad(2\le i\le m).
\]
The pairs among $1,2,3$, and every pair containing vertex $1$, satisfy $\eta_{ij}=v_i^{(g)}+v_j^{(g)}$ by construction. This completes the case $m=3$.

When $m\ge4$, take four distinct vertices $i,j,k,l\in G_g$. Giving each within-group degree one and setting every other count to zero leaves exactly the three perfect matchings. Their equal weights imply
\begin{equation}
 \eta_{ij}+\eta_{kl}=\eta_{ik}+\eta_{jl}=\eta_{il}+\eta_{jk}.
 \label{eq:odds-matching}
\end{equation}
Applying \eqref{eq:odds-matching} to $1,2,3,k$ for $k\ge4$ gives $\eta_{2k}=v_2^{(g)}+v_k^{(g)}$ and $\eta_{3k}=v_3^{(g)}+v_k^{(g)}$. Applying it to $1,2,k,l$ for $4\le k<l\le m$ gives the remaining pairs. Thus the within-group representation holds for every group size.

Returning to the original vertex labels, set $\gamma_{i,h}=r_i^{(gh)}$ for $i\in G_g$ and $\gamma_{j,g}=c_j^{(gh)}$ for $j\in G_h$, for each $g<h$. Set $\gamma_{i,g}=v_i^{(g)}$ within $G_g$. Each coefficient slot $(i,h)$ belongs to the unique block $\{g(i),h\}$: it receives a cross-group coefficient when $h\ne g(i)$ and a within-group coefficient when $h=g(i)$. Hence no slot receives conflicting assignments, establishing (iii).

Finally, under (iii), every $a\in\mathcal F(y,B)$ satisfies
\begin{align*}
 \sum_{i<j}\eta_{ij}a_{ij}
 &=\sum_{i<j}a_{ij}\{\gamma_{i,g(j)}+\gamma_{j,g(i)}\}\\
 &=\sum_i\sum_{h=1}^{K}\gamma_{ih}\sum_{j\in G_h}a_{ij}
 =\sum_i\sum_{h=1}^{K}\gamma_{ih}y_{ih}.
\end{align*}
Thus the conditional weights are constant on the fibre, proving (ii).
\end{proof}

For the independent Bernoulli model, the symmetric divergence in \eqref{eq:symmetric-kl} vanishes on any fibre where the two sets of edge marginals agree. The conditional graph law is then uniform on that fibre. A singleton fibre, for example, is uniform for any strictly interior $P^\star$. To obtain the additive log odds in \eqref{eq:block-additive-odds}, the proof uses the same $P^\star$ across fibres. Comparing the two matchings across groups and the three perfect matchings on four vertices within a group suffices to obtain all the odds restrictions, for any group sizes. The compatible graphs then have equal conditional probability, even when the generating edge probabilities and the conditional edge marginals vary across pairs.

In both this result and \citet[Lemma~A.1]{tseng2026residual}, conditioning on counts cancels additive node effects. The lemma fixes the edge between two respondents and each respondent's number of other contacts. Their own activity terms then cancel, while the weights of their possible contacts remain. Proposition~\ref{prop:conditional-uniform} conditions on all roster counts and characterizes when the complete conditional graph law is uniform on every feasible fibre.

\paragraph{Aligned blockmodels.}
In a stochastic blockmodel (SBM), edges are conditionally independent given the generating-community labels, and their probabilities depend only on the communities of their two endpoints \citep{holland1983}. These generating communities need not coincide with the observation groups used to collect ARD. Let $C_i$ denote the generating-community label of node $i$. 
Suppose $p^\star_{ij}=\theta_{C_iC_j}$ for a symmetric matrix with entries strictly between zero and one, and the observation partition refines the generating communities. Write $C_i=c(g(i))$. Taking
\[
 \gamma_{ih}=\tfrac12\log\frac{\theta_{c(g(i)),c(h)}}
                                      {1-\theta_{c(g(i)),c(h)}}
\]
satisfies \eqref{eq:block-additive-odds}, so the proposition gives the aligned-SBM equality. Equivalently, each observation block has constant odds and a fixed edge count,
\begin{equation}
 m_{gh}=\sum_{i\in G_g}y_{ih}\quad(g<h),\qquad
 m_{gg}=\tfrac12\sum_{i\in G_g}y_{ig}.
 \label{eq:sbm-aligned-counts}
\end{equation}
Its log-weight is therefore constant on the feasible graphs.

Alignment is sufficient but not necessary. For example, an SBM with $\operatorname{logit}\theta_{rs}=\beta_r+\beta_s$ satisfies \eqref{eq:block-additive-odds} for any observation partition by taking $\gamma_{ih}=\beta_{C_i}$, including when observation groups mix generating communities.

A parameter mixture supported entirely on the class in Proposition~\ref{prop:conditional-uniform} retains the same uniform conditional law, since every component has that law on $\mathcal F(y,B)$. In particular, uncertainty over community assignments that all coarsen the fixed observation partition does not invalidate the equality. General mixtures over unrestricted communities can include models outside this class and can couple blocks. The mixed-community design in Section~\ref{sec:sbm} permits residual odds variation, but does not guarantee an oracle--uniform gap in every block or realization.

%% file: plugin-method.tex
\subsection{Conditioning the fitted matrices}\label{sec:plugin-method}
We call the entries of a fitted probability matrix its \emph{direct scores}. To separate direct scoring from conditioning, we insert each saved matrix into the weighted law in \eqref{eq:weighted} and call the resulting conditional edge marginals \emph{plug-in scores}. For an interior matrix $\widetilde P$, define
\begin{equation}
 \widehat q(a\mid Y,B)\propto
 \mathbf1\{a\in\mathcal F(Y,B)\}
 \prod_{i<j}\left(\frac{\widetilde p_{ij}}{1-\widetilde p_{ij}}\right)^{a_{ij}},
 \qquad Q^{\mathrm{plug}}=\E_{\widehat q}[A].
 \label{eq:plugin}
\end{equation}
This construction uses only the observed counts, memberships and fitted matrix. Once that matrix is fixed, the same block enumeration gives $Q^{\mathrm{plug}}B=Y$ and the same exact forced states as the references.

For circular, $\widetilde P$ is the saved mean probability matrix over the four chains. We condition this mean matrix once. Averaging conditional marginals over individual parameter draws would define a different procedure. For nuclear, exact zero and one entries can assign zero probability to the observed count array. We therefore set $\widetilde p_{ij}=\min(1-\epsilon,\max(\epsilon,\widehat p_{ij}))$ for $i<j$, with fixed $\epsilon=10^{-6}$, and repeat at $10^{-4}$ and $10^{-8}$. The diagonal remains zero. We retain a clipping-only score to distinguish this regularization from conditioning. Equation~\eqref{eq:plugin} treats the matrix estimated from $Y$ as fixed. A posterior over graphs would also integrate uncertainty in the generating probabilities.

We specified this procedure and its clipping thresholds after the direct-score comparisons and before calculating the plug-in scores. We used the saved fits for forty exploratory graphs, forty fresh replicate blocks each evaluated at three circular densities, and forty separate blockmodel graphs, all at $K=8$.

%% file: evaluation.tex
Let \(\mathcal E_1,\mathcal E_0\) be the unordered edge and nonedge sets in a chosen evaluation domain. We use
\[
 \operatorname{AUC}(T,A)=
 \frac{\sum_{e\in\mathcal E_1}\sum_{f\in\mathcal E_0}
 [\mathbf1\{T_e>T_f\}+\tfrac12\mathbf1\{T_e=T_f\}]}
 {|\mathcal E_1||\mathcal E_0|}.
\]
On the full set of unordered pairs, raw AUC evaluates the ranking by $T$ and zero-adjusted AUC evaluates the ranking by the zero-adjusted score $T^Z$. To measure ranking among the pairs that remain after zero exclusion or after applying the complete constraints, we also evaluate $T$ within $\mathcal U$ and $\mathcal V$, respectively. The conditional references and the evaluation on $\mathcal V$ are available only at $K=8$. Each domain contains both edges and nonedges in every graph evaluated on that domain.

Generating-probability RMSE and count RMSE are
\[
 \frac{\|\widehat P-P^\star\|_F}{\sqrt{n(n-1)}},
 \qquad
 \frac{\|\widehat P B-Y\|_F}{\sqrt{nK}},
\]
respectively. For conditional graph prediction we report realized Brier scores and the conditional excess risk in \eqref{eq:brier}. We use the count product only for ranking.

The exploratory and blockmodel samples each contain forty independent graphs. The fresh density study contains forty independent blocks, each contributing one graph at each of three coupled densities. The Monte Carlo standard error (MCSE) of a mean is the sample standard deviation across graphs or blocks divided by \(\sqrt{40}\). For method comparisons, we first compute the difference within each graph. These MCSEs quantify sampling variation across replicates, separately from Markov chain Monte Carlo (MCMC) error in a fitted matrix. All fits and chains contribute to the summaries, including the final nuclear iterates from runs that reached their iteration limit (Appendices~\ref{app:exploratory}--\ref{app:independent-checks}).

Separate implementations checked the conditional marginals and comparisons against saved simulation outputs. The reported graph statistics, estimated probabilities and ranking scores were computed by those programs from the simulated graphs and model inputs.

%% file: density-study.tex
\subsubsection{Design and protocol}
The fresh circular study evaluates the prespecified comparisons in forty independent simulation blocks, separate from the exploratory sample (Section~\ref{sec:numerical}). Within a block, the three densities share the same independent uniform latent angles for $n=40$ nodes and the same independent uniforms for the 780 unordered pairs. We observe each graph under two nested balanced partitions, giving 120 graphs and 240 graph--partition cases. With nodes indexed from zero, $g_K(i)=i\bmod K$ for $K=4,8$. Every node reports its counts.

The interaction strength is fixed at $\zeta=1.5$. We retain the exploratory design's intercept of $-2.8$ for the fresh sparse sample, giving a population mean edge probability of approximately $0.085279$ (8.53\%). For target means of $0.15$ and $0.25$, deterministic quadrature and root finding before simulation gave intercepts of approximately $-2.103$ and $-1.370$ (Appendix~\ref{app:density-checks}). Generation and fitting use separate random-number streams.

Each circular fit averages the 12,800 probability matrices from four chains, each with 1,600 warmup and 3,200 retained sweeps. Appendix~\ref{app:exploratory-protocol} specifies the prior, working likelihood, starts, proposals and warmup adaptation. The nuclear penalties are $\lambda\in\{0.25,1,4\}$. For the fresh and blockmodel studies we use the alternating direction method of multipliers (ADMM) with residual balancing. Before generating the fresh graphs, we fixed the update schedule, residual tolerance of $10^{-7}$ and cap of 200,000 iterations (Appendix~\ref{app:density-checks}).

We compute the conditional references for all 120 fresh graphs at $K=8$. For comparisons across densities, we compute each change in a score or method difference within a block before averaging over the forty blocks.

\subsubsection{Constraint accounting and residual ranking}
In the sparse eight-group design, the complete counts fix an average of 35.6\% of realized edges. Zero counts alone exclude 91.1\% of nonedges, compared with 91.7\% under the complete constraints (Table~\ref{tab:fresh-design}). At 15\% and 25\% density, the fraction of realized edges fixed by the counts falls to 22.8\% and 14.3\%, respectively. The mean number of unresolved pairs rises from 102.65 in the sparse setting to 230.47 and 428.05. A score that assigns every unresolved pair the same intermediate rank attains mean full-pair AUCs of 0.973, 0.918 and 0.809 at the three densities.

\begin{table}[!ht]
 \centering\small
 \setlength{\tabcolsep}{3.5pt}
 \input{tables/density-design.tex}
 \caption{Constraint accounting in the fresh sample, mean (MCSE) across forty blocks per setting. Column headings give population mean edge probabilities. Percentages use all nonedges or all realized edges as the respective denominators, with MCSEs in percentage points. The forced-state-only score assigns $0$, $1/2$ and $1$ to forced zeros, unresolved pairs and forced ones, respectively. Complete-constraint quantities are available only at $K=8$.}
 \label{tab:fresh-design}
\end{table}

\begin{figure}[!ht]
 \centering
 \includegraphics[width=\linewidth]{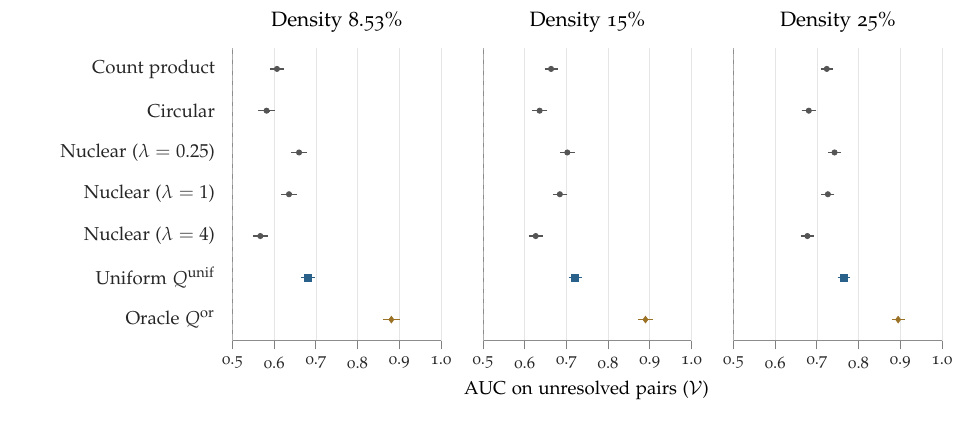}
 \caption{Ranking on the unresolved pairs in the fresh circular study at $K=8$. Points show mean AUC and bars extend one MCSE in each direction. The same forty independent blocks contribute at each density. The uniform reference weights compatible graphs equally, and the oracle uses the generating odds. The fitted scores use matrix entries directly, including all three nuclear penalties. The circular fit ranks below uniform here and above uniform at each density when its odds weight compatible graphs (Table~\ref{tab:plugin}). Table~\ref{tab:fresh-ranking} gives paired differences.}
 \label{fig:density-unresolved}
\end{figure}

On $\mathcal V$, the uniform reference ranks edges more accurately on average than any direct fitted score at each density (Figure~\ref{fig:density-unresolved} and Table~\ref{tab:fresh-ranking}). In the sparse setting, mean AUC is 0.681 for uniform, 0.636 for nuclear at $\lambda=1$ and 0.582 for circular. We designated $\lambda=1$ as the main nuclear reference before generating the fresh sample. In order of increasing density, the paired mean AUC advantages of uniform over this fit are $0.045$, $0.036$ and $0.038$. Of the three fixed nuclear penalties, $\lambda=0.25$ gives the highest observed mean direct AUC at each density, trailing uniform by $0.021$, $0.019$ and $0.022$. Weighting the same feasible graphs by the generating odds raises mean AUC above uniform by $0.200$, $0.169$ and $0.130$, respectively.

\begin{table}[!ht]
 \centering\small
 \setlength{\tabcolsep}{3.5pt}
 \input{tables/density-ranking.tex}
 \caption{Ranking within $\mathcal V$ at $K=8$ in the fresh sample, mean (MCSE) across forty blocks. The lower panel reports within-block differences. $Q^{\mathrm{unif}}$ assigns equal weight to feasible graphs, and $Q^{\mathrm{or}}$ weights them by the generating odds.}
 \label{tab:fresh-ranking}
\end{table}

The unresolved set changes with density. The rising AUCs within $\mathcal V$ and the falling full-pair AUC of the forced-state-only score reflect different sets of edge--nonedge comparisons.

\FloatBarrier
\subsubsection{Probability error and the count-product comparison}
After dropping every pair ruled out by a zero count, does the fitted nuclear matrix rank edges better than multiplying the two reported counts? The surviving pairs form $\mathcal U$. Some of their states are fixed by the remaining count constraints. On this set, the AUC difference between the nuclear fit at $\lambda=1$ and the count product varies with both partition and density (Table~\ref{tab:fresh-estimation}). At $K=4$, nuclear minus count-product AUC changes from $+0.009$ at the sparse density to $-0.013$ and $-0.031$. At 25\% density it is negative in all forty blocks. At $K=8$ the mean difference remains positive but contracts from $+0.062$ to $+0.040$ and $+0.015$.

\begin{table}[!ht]
 \centering\small
 \setlength{\tabcolsep}{3pt}
 \input{tables/density-estimation.tex}
 \caption{Generating-probability error and paired ranking differences in the fresh sample, mean (MCSE) across forty blocks. RMSE is measured against $P^\star$. The last row for each partition gives nuclear minus count-product AUC within $\mathcal U$. All nuclear entries use $\lambda=1$. The reduction rows give $100\{1-\overline{\mathrm{RMSE}}_{\rm Circular}/\overline{\mathrm{RMSE}}_{\rm Density}\}$ from unrounded means, without an MCSE.}
 \label{tab:fresh-estimation}
\end{table}

The circular fit reduces mean generating-probability RMSE relative to the density baseline by 1.97\%, 4.91\% and 9.20\% at $K=8$, and by 1.16\%, 2.44\% and 3.85\% at $K=4$. The nuclear fit at $\lambda=1$ has higher mean RMSE than the baseline at both partitions and all three densities. At $K=8$ and 25\% density, for example, nuclear RMSEs are 0.276, 0.250 and 0.211 for $\lambda=0.25,1,4$, compared with 0.174 for density and 0.158 for circular.

\paragraph{Numerical checks.}
All 240 circular fits pass the diagnostic screens. Of 720 nuclear optimization runs, 719 meet both residual tolerances, including every $K=8$ run. One $K=4$, $\lambda=4$ run reaches the iteration cap with primal residual $2.27\times10^{-7}$, above the $10^{-7}$ threshold. Its dual residual passes. The final feasible matrix contributes to the comparison as a capped fit (Appendix~\ref{app:density-checks}).

%% file: tables/density-design.tex
\begin{tabular}{llrrr}
\toprule
$K$ & Diagnostic & 8.53\% & 15\% & 25\% \\
\midrule
4 & Zero-excluded nonedges (\%) & 69.7 (0.8) & 41.7 (1.1) & 14.9 (0.7) \\
4 & Nonexcluded pairs $|\mathcal U|$ & 283.07 (6.28) & 503.52 (7.49) & 693.10 (4.48) \\
\midrule
8 & Zero-excluded nonedges (\%) & 91.1 (0.3) & 77.6 (0.6) & 53.2 (0.9) \\
8 & Forced nonedges (\%) & 91.7 (0.3) & 79.0 (0.6) & 55.5 (0.9) \\
8 & Nonexcluded pairs $|\mathcal U|$ & 130.40 (3.27) & 266.62 (5.24) & 469.35 (6.27) \\
8 & Unresolved pairs $|\mathcal V|$ & 102.65 (3.36) & 230.47 (5.79) & 428.05 (6.93) \\
8 & Forced present edges (\%) & 35.6 (1.3) & 22.8 (1.1) & 14.3 (0.8) \\
8 & Forced-state-only AUC & 0.973 (0.0014) & 0.918 (0.0033) & 0.809 (0.0051) \\
\bottomrule
\end{tabular}

%% file: tables/density-ranking.tex
\begin{tabular}{lrrr}
\toprule
Score; AUC on $\mathcal V$ & 8.53\% & 15\% & 25\% \\
\midrule
Count product & 0.607 (0.0044) & 0.663 (0.0030) & 0.724 (0.0021) \\
Circular & 0.582 (0.0071) & 0.636 (0.0059) & 0.681 (0.0050) \\
Nuclear ($\lambda=0.25$) & 0.660 (0.0071) & 0.702 (0.0050) & 0.742 (0.0029) \\
Nuclear ($\lambda=1$) & 0.636 (0.0060) & 0.684 (0.0048) & 0.726 (0.0032) \\
Nuclear ($\lambda=4$) & 0.567 (0.0055) & 0.626 (0.0048) & 0.677 (0.0037) \\
$Q^{\mathrm{unif}}$ & 0.681 (0.0043) & 0.721 (0.0033) & 0.764 (0.0023) \\
$Q^{\mathrm{or}}$ & 0.881 (0.0082) & 0.890 (0.0049) & 0.895 (0.0029) \\
\midrule
Paired AUC difference &  &  &  \\
\midrule
$Q^{\mathrm{unif}} - \text{Nuclear}$ ($\lambda=0.25$) & $+0.0210$ (0.0057) & $+0.0186$ (0.0033) & $+0.0221$ (0.0021) \\
$Q^{\mathrm{unif}} - \text{Nuclear}$ ($\lambda=1$) & $+0.0451$ (0.0059) & $+0.0365$ (0.0033) & $+0.0380$ (0.0024) \\
$Q^{\mathrm{unif}} - \text{Circular}$ & $+0.0989$ (0.0079) & $+0.0850$ (0.0060) & $+0.0838$ (0.0053) \\
$Q^{\mathrm{or}}-Q^{\mathrm{unif}}$ & $+0.2003$ (0.0097) & $+0.1689$ (0.0058) & $+0.1305$ (0.0032) \\
\bottomrule
\end{tabular}

%% file: tables/density-estimation.tex
\begin{tabular}{llrrr}
\toprule
$K$ & Estimator or comparison & 8.53\% & 15\% & 25\% \\
\midrule
4 & Density RMSE & 0.0712 (0.00026) & 0.1171 (0.00034) & 0.1740 (0.00044) \\
4 & Circular RMSE & 0.0703 (0.00031) & 0.1143 (0.00053) & 0.1673 (0.00084) \\
4 & Circular RMSE reduction (\%) & 1.16 & 2.44 & 3.85 \\
4 & Nuclear ($\lambda=1$) RMSE & 0.1238 (0.00129) & 0.1802 (0.00155) & 0.2459 (0.00124) \\
4 & $\text{Nuclear} - \text{product}$ AUC$_{\mathcal U}$ & $+0.0087$ (0.0039) & $-0.0129$ (0.0025) & $-0.0310$ (0.0023) \\
\midrule
8 & Density RMSE & 0.0712 (0.00026) & 0.1171 (0.00034) & 0.1740 (0.00044) \\
8 & Circular RMSE & 0.0698 (0.00038) & 0.1114 (0.00067) & 0.1580 (0.00121) \\
8 & Circular RMSE reduction (\%) & 1.97 & 4.91 & 9.20 \\
8 & Nuclear ($\lambda=1$) RMSE & 0.1325 (0.00140) & 0.1889 (0.00153) & 0.2499 (0.00122) \\
8 & $\text{Nuclear} - \text{product}$ AUC$_{\mathcal U}$ & $+0.0615$ (0.0046) & $+0.0400$ (0.0035) & $+0.0150$ (0.0024) \\
\bottomrule
\end{tabular}

%% file: sbm-study.tex
\subsubsection{Design and matching}
The second generating family is an undirected Bernoulli stochastic blockmodel \citep{holland1983}. Each of forty independent graphs has $n=40$ nodes in four generating communities of ten. For each graph, we draw the community labels $C$ by a uniform permutation of this balanced allocation, independently of the fixed balanced $K=8$ observation partition $B$ and the edge-generation uniforms. We retain every allocation regardless of its overlap with $B$. Given $C$, unordered edges are independent, with probabilities $p_{\mathrm{in}}$ within communities and $p_{\mathrm{out}}$ between them. We observe $Y=AB$.

The parameters match the population dyad mean and probability variance of the circular model at $\alpha\approx-2.103$ and $\zeta=1.5$. With within-community dyad fraction $\omega=180/780=3/13$, target mean $\mu_P=0.15$ and variance $v\approx0.01357$, set
\begin{equation}
 \Delta=\sqrt{\frac{v}{\omega(1-\omega)}},\qquad
 p_{\mathrm{in}}=\mu_P+(1-\omega)\Delta,\qquad
 p_{\mathrm{out}}=\mu_P-\omega\Delta.
 \label{eq:sbm-match}
\end{equation}
The resulting probabilities, $p_{\mathrm{in}}\approx0.3627$ and $p_{\mathrm{out}}\approx0.0862$, were fixed before graph generation (Appendix~\ref{app:sbm-checks}). The two families also have the same average Bernoulli noise, $\mu_P(1-\mu_P)-v$. The balanced blockmodel gives every node the same expected degree. Expected degrees in a finite circular configuration can vary, and the matched moments allow the difficulty of conditional reconstruction to differ.

Both fits use the settings in Section~\ref{sec:fresh}, including all three nuclear penalties. The circular fit is misspecified here. True communities and $P^\star$ enter only graph generation, the oracle and evaluation. The primary comparisons concern the fractions of fixed pairs, the AUC differences between uniform and count product and between oracle and uniform, and the conditional excess Brier risk of uniform. Direct fitted scores give a secondary comparison. Methods are paired within each SBM graph. Comparisons with the independent circular sample are unpaired.

\subsubsection{Conditional references and fitted scores}
The complete counts fix an average of 22.2\% of realized edges and 78.5\% of nonedges in the SBM, leaving 234.38 of 780 pairs unresolved (Table~\ref{tab:sbm-context}). In the matched 15\% circular sample the counts fix 22.8\% of edges and 79.0\% of nonedges, leaving 230.47 pairs unresolved on average.

\begin{table}[htbp]
 \centering\small
 \setlength{\tabcolsep}{3pt}
 \input{tables/sbm-context.tex}
 \caption{Constraint accounting and within-SBM AUC contrasts, mean (MCSE). The upper panel compares two independent samples of forty graphs each, the circular 15\% setting and the SBM. Percentages use the indicated class denominators, with MCSEs in percentage points. The lower panel averages within-graph differences over the forty SBM graphs.}
 \label{tab:sbm-context}
\end{table}

Within $\mathcal V$, uniform AUC exceeds count-product AUC by $0.056$ on average (MCSE $0.0015$), and oracle AUC exceeds uniform AUC by $0.146$ ($0.0060$). Both differences are positive in all forty graphs. Uniform has mean conditional excess Brier risk $0.065$ ($0.0020$), the mean squared distance from the conditional oracle in \eqref{eq:brier} (Table~\ref{tab:sbm-comparison}).

\begin{table}[htbp]
 \centering\small
 \setlength{\tabcolsep}{3pt}
 \input{tables/sbm-comparison.tex}
 \caption{SBM ranking and probability errors, mean (MCSE) across forty independent graphs. Raw AUC uses all unordered pairs. RMSE targets the generating $P^\star$, and conditional excess Brier risk targets reconstruction of the realized graph within $\mathcal V$. Dashes mark inapplicable targets for the conditional references and the count product.}
 \label{tab:sbm-comparison}
\end{table}

On the unresolved pairs, scoring by $P^\star$ gives only two possible scores in this blockmodel, one for within-community pairs and one for between-community pairs. Its mean AUC is $0.697$, below uniform at $0.729$. Conditioning those probabilities on the observed counts raises mean AUC to $0.875$ on the same pairs and from $0.692$ to $0.979$ on all pairs (Table~\ref{tab:sbm-comparison}). Direct fitted scores also rank below uniform on $\mathcal V$, with mean AUC $0.619$ for circular and $0.706$, $0.688$ and $0.631$ for nuclear at $\lambda=0.25,1,4$. Against $P^\star$, the circular fit has mean RMSE $0.114$, compared with $0.117$ for the density baseline. Each nuclear fit has higher mean RMSE than the baseline.

\paragraph{Numerical checks.}
All 120 nuclear fits meet the residual tolerances, and all forty four-chain circular fits pass the diagnostic screens. Independent calculations verify the conditional references (Appendix~\ref{app:sbm-checks}).

%% file: tables/sbm-context.tex
\begin{tabular}{lrr}
\toprule
Constraint accounting & Circular, 15\% & SBM \\
\midrule
Zero-excluded nonedges (\%) & 77.6 (0.6) & 77.2 (0.6) \\
Complete-constraint forced nonedges (\%) & 79.0 (0.6) & 78.5 (0.5) \\
Complete-constraint forced edges (\%) & 22.8 (1.1) & 22.2 (1.1) \\
Unresolved pairs $|\mathcal V|$ & 230.47 (5.79) & 234.38 (5.11) \\
\midrule
\multicolumn{2}{l}{Within-SBM paired AUC$_{\mathcal V}$ difference} & Mean (MCSE) \\
\midrule
\multicolumn{2}{l}{$Q^{\mathrm{unif}} - \text{count product}$} & $+0.0563$ (0.0015) \\
\multicolumn{2}{l}{$Q^{\mathrm{or}}-Q^{\mathrm{unif}}$} & $+0.1464$ (0.0060) \\
\multicolumn{2}{l}{$Q^{\mathrm{unif}} - \text{Nuclear}$ ($\lambda=0.25$)} & $+0.0227$ (0.0041) \\
\multicolumn{2}{l}{$Q^{\mathrm{unif}} - \text{Nuclear}$ ($\lambda=1$)} & $+0.0408$ (0.0038) \\
\bottomrule
\end{tabular}

%% file: tables/sbm-comparison.tex
\begin{tabular}{lrrrr}
\toprule
Score or estimator & Raw AUC & AUC$_{\mathcal V}$ & $P^\star$ RMSE & \shortstack{Conditional excess\\Brier on $\mathcal V$} \\
\midrule
Count product & 0.932 (0.0017) & 0.672 (0.0029) & --- & --- \\
Circular working fit & 0.820 (0.0040) & 0.619 (0.0050) & 0.1135 (0.00056) & 0.1467 (0.0027) \\
Nuclear ($\lambda=0.25$) & 0.945 (0.0017) & 0.706 (0.0041) & 0.2250 (0.00195) & 0.0725 (0.0019) \\
Nuclear ($\lambda=1$) & 0.927 (0.0018) & 0.688 (0.0039) & 0.1898 (0.00185) & 0.0831 (0.0021) \\
Nuclear ($\lambda=4$) & 0.864 (0.0029) & 0.631 (0.0046) & 0.1479 (0.00089) & 0.1209 (0.0026) \\
Uniform $Q^{\mathrm{unif}}$ & 0.954 (0.0015) & 0.729 (0.0029) & --- & 0.0649 (0.0020) \\
Oracle $Q^{\mathrm{or}}$ & 0.979 (0.0011) & 0.875 (0.0057) & --- & 0.0000 (0.0000) \\
True $P^\star$ & 0.692 (0.0044) & 0.697 (0.0056) & 0.0000 (0.00000) & 0.1157 (0.0020) \\
Constant density & 0.500 (0.0000) & 0.500 (0.0000) & 0.1173 (0.00015) & 0.1620 (0.0029) \\
\bottomrule
\end{tabular}

%% file: exploratory-summary.tex
The comparison originated in the sparse circular sample, which combined the four graphs of \citet{tseng2025comparison} with thirty-six subsequent graphs. We developed the count diagnostics, conditional references and stability checks after inspecting its initial results. Appendix~\ref{app:exploratory} gives the exploratory protocol, probability errors, penalty sensitivity and numerical checks.

At $K=8$, the complete counts fix an average of 24.95 present edges and 655.65 nonedges, leaving 99.4 pairs unresolved. The count product attains mean AUC 0.970 over all pairs and 0.613 on the unresolved pairs (Table~\ref{tab:ranking}). On the unresolved pairs, uniform has mean AUC 0.680, compared with 0.643 for nuclear at $\lambda=1$ and 0.594 for circular. Here ranking pairs directly by the generating probabilities gives higher mean AUC than uniform weighting (0.747 versus 0.680), reversing the order observed in the blockmodel (Table~\ref{tab:sbm-comparison}). Using both the generating odds and the observed counts gives the highest mean AUC in both samples (0.889 here and 0.875 in the blockmodel).

\begin{table}[htbp]
 \centering\small
 \setlength{\tabcolsep}{3.5pt}
 \input{tables/ranking.tex}
 \caption{Ranking at \(K=8\), mean (MCSE) across the forty exploratory graphs. In the raw AUC column, \(P^\star\) scores pairs before the counts are observed, while the count product and fitted scores use counts from the realized graph. The oracle \(Q^{\mathrm{or}}\) uses both \(P^\star\) and the counts. Zero-adjusted AUC evaluates \(T^Z\) over all pairs. The sets \(\mathcal U\) and \(\mathcal V\) retain pairs that survive zero exclusion and pairs unresolved by the complete constraints, respectively. Nuclear uses \(\lambda=1\).}
 \label{tab:ranking}
\end{table}

The ordering of circular and nuclear direct scores depends on the penalty. All three nuclear fits have lower mean unresolved-pair AUC than uniform. All eighty circular fits pass the numerical screens. Five of the 240 exploratory nuclear fits reach the iteration cap and remain in the comparisons (Appendix~\ref{app:exploratory-checks}).

%% file: tables/ranking.tex
\begin{tabular}{lrrrr}
\toprule
Score & Raw AUC & Zero-adjusted AUC & AUC on $\mathcal{U}$ & AUC on $\mathcal{V}$ \\
\midrule
Count product & 0.970 (0.0013) & 0.970 (0.0013) & 0.654 (0.0058) & 0.613 (0.0056) \\
True $P^\star$ & 0.750 (0.0049) & 0.978 (0.0010) & 0.746 (0.0086) & 0.747 (0.0100) \\
Circular & 0.901 (0.0039) & 0.969 (0.0015) & 0.633 (0.0070) & 0.594 (0.0067) \\
Nuclear ($\lambda=1$) & 0.967 (0.0014) & 0.976 (0.0012) & 0.715 (0.0074) & 0.643 (0.0066) \\
$Q^{\mathrm{unif}}$ & 0.984 (0.0010) & 0.984 (0.0010) & 0.811 (0.0066) & 0.680 (0.0052) \\
$Q^{\mathrm{or}}$ & 0.995 (0.0005) & 0.995 (0.0005) & 0.937 (0.0050) & 0.889 (0.0097) \\
\bottomrule
\end{tabular}

%% file: plugin-study.tex
Conditioning the same saved circular matrices reverses their mean AUC comparison with uniform (Table~\ref{tab:plugin}). In this post hoc analysis, the circular fits to circular data use the generating form for the edge probabilities, with interaction strength fixed at its true value $\zeta=1.5$. In order of increasing density in the coupled fresh sample, mean AUC on $\mathcal V$ rises from $0.582$ to $0.696$, from $0.636$ to $0.740$ and from $0.681$ to $0.782$. The paired mean gains over uniform are $0.0159$ (MCSE $0.0044$), $0.0191$ ($0.0027$) and $0.0177$ ($0.0020$), smaller at each density than the gain from using the generating odds (Table~\ref{tab:fresh-ranking}).

\begin{table}[!ht]
 \centering\small
 \setlength{\tabcolsep}{3pt}
 \renewcommand{\arraystretch}{0.96}
 \input{tables/plugin-comparison.tex}
 \caption{Direct and plug-in scores on the same unresolved set $\mathcal V$ at $K=8$. The three fresh circular panels share forty blocks. The SBM and exploratory panels each summarize a separate sample of forty graphs. Panel headings give the uniform reference values on $\mathcal V$. The first two numeric columns report mean AUC, with the MCSE of the plug-in mean in parentheses. The last two columns report plug-in minus uniform, with the MCSE of each paired difference computed from unrounded graph-level values. Positive AUC differences and negative excess Brier differences favor the plug-in score. Nuclear uses $\epsilon=10^{-6}$ and includes the capped exploratory fit. Circular uses its saved mean matrix without clipping.}
 \label{tab:plugin}
\end{table}

In the separate SBM sample, where the circular model is misspecified, conditioning raises mean AUC from $0.619$ to $0.736$, a gain over uniform of $0.0073$ (MCSE $0.0024$). In the original exploratory circular sample, conditioning raises mean AUC from $0.594$ to $0.696$, compared with $0.680$ for uniform. The circular plug-in scores also have lower mean conditional excess Brier risk than uniform at the three fresh densities and in the SBM and exploratory samples.

At all three penalties, the nuclear plug-in scores have lower mean AUC and higher mean conditional excess Brier risk than uniform in each panel of Table~\ref{tab:plugin}. The change from direct to plug-in AUC depends on the penalty and density. Conditioning the $\lambda=1$ fit raises mean AUC at the two lower fresh densities and lowers it at 25\%. At $\lambda=0.25$, conditioning lowers mean AUC in each panel. Every plug-in matrix reproduces the counts, but its ranking on $\mathcal V$ depends on the fitted odds used to weight the feasible graphs.

\paragraph{Numerical checks.}
The plug-in calculation reproduces the original feasible supports and reference marginals. The clipping sensitivity and clipping-only comparisons preserve the reported mean ordering. Appendix~\ref{app:plugin-checks} reports the numerical errors in the count constraints and reference marginals, along with the tie checks.

%% file: tables/plugin-comparison.tex
\begin{tabular}{lrrrr}
\toprule
Saved fit & Direct AUC & Plug-in AUC & \shortstack{AUC difference\\from uniform} & \shortstack{Excess Brier difference\\from uniform} \\
\midrule
\multicolumn{5}{l}{\textit{Fresh circular, 8.53\%}; uniform AUC 0.681; uniform excess Brier 0.0865} \\
Circular & 0.582 & 0.696 (0.0056) & $+0.0159$ (0.0044) & $-0.0017$ (0.0008) \\
Nuclear ($\lambda=0.25$) & 0.660 & 0.653 (0.0114) & $-0.0276$ (0.0106) & $+0.0113$ (0.0018) \\
Nuclear ($\lambda=1$) & 0.636 & 0.668 (0.0096) & $-0.0126$ (0.0098) & $+0.0070$ (0.0017) \\
Nuclear ($\lambda=4$) & 0.567 & 0.664 (0.0080) & $-0.0163$ (0.0087) & $+0.0080$ (0.0017) \\
\midrule
\multicolumn{5}{l}{\textit{Fresh circular, 15\%}; uniform AUC 0.721; uniform excess Brier 0.0736} \\
Circular & 0.636 & 0.740 (0.0040) & $+0.0191$ (0.0027) & $-0.0042$ (0.0007) \\
Nuclear ($\lambda=0.25$) & 0.702 & 0.687 (0.0068) & $-0.0338$ (0.0054) & $+0.0142$ (0.0017) \\
Nuclear ($\lambda=1$) & 0.684 & 0.691 (0.0059) & $-0.0299$ (0.0046) & $+0.0112$ (0.0013) \\
Nuclear ($\lambda=4$) & 0.626 & 0.689 (0.0058) & $-0.0320$ (0.0042) & $+0.0125$ (0.0012) \\
\midrule
\multicolumn{5}{l}{\textit{Fresh circular, 25\%}; uniform AUC 0.764; uniform excess Brier 0.0626} \\
Circular & 0.681 & 0.782 (0.0027) & $+0.0177$ (0.0020) & $-0.0067$ (0.0007) \\
Nuclear ($\lambda=0.25$) & 0.742 & 0.723 (0.0036) & $-0.0419$ (0.0030) & $+0.0183$ (0.0010) \\
Nuclear ($\lambda=1$) & 0.726 & 0.721 (0.0040) & $-0.0438$ (0.0031) & $+0.0189$ (0.0011) \\
Nuclear ($\lambda=4$) & 0.677 & 0.718 (0.0043) & $-0.0468$ (0.0034) & $+0.0185$ (0.0011) \\
\midrule
\multicolumn{5}{l}{\textit{Stochastic blockmodel}; uniform AUC 0.729; uniform excess Brier 0.0649} \\
Circular & 0.619 & 0.736 (0.0033) & $+0.0073$ (0.0024) & $-0.0017$ (0.0006) \\
Nuclear ($\lambda=0.25$) & 0.706 & 0.687 (0.0059) & $-0.0415$ (0.0061) & $+0.0151$ (0.0016) \\
Nuclear ($\lambda=1$) & 0.688 & 0.695 (0.0054) & $-0.0335$ (0.0054) & $+0.0124$ (0.0014) \\
Nuclear ($\lambda=4$) & 0.631 & 0.696 (0.0051) & $-0.0328$ (0.0047) & $+0.0122$ (0.0012) \\
\midrule
\multicolumn{5}{l}{\textit{Exploratory circular}; uniform AUC 0.680; uniform excess Brier 0.0887} \\
Circular & 0.594 & 0.696 (0.0070) & $+0.0152$ (0.0048) & $-0.0021$ (0.0008) \\
Nuclear ($\lambda=0.25$) & 0.665 & 0.656 (0.0095) & $-0.0242$ (0.0084) & $+0.0109$ (0.0021) \\
Nuclear ($\lambda=1$) & 0.643 & 0.662 (0.0088) & $-0.0187$ (0.0066) & $+0.0074$ (0.0016) \\
Nuclear ($\lambda=4$) & 0.578 & 0.662 (0.0094) & $-0.0180$ (0.0080) & $+0.0091$ (0.0019) \\
\bottomrule
\end{tabular}

%% file: experiment.tex
\subsection{Data and computational protocol}\label{app:exploratory-protocol}
The exploratory sample comprises four graphs from the matched-input study of \citet{tseng2025comparison} (seeds 101, 202, 303 and 404) and thirty-six additional graphs (seeds 1001--1036). Each of the forty independent graphs has \(n=40\) nodes, independent uniform latent angles on \([-\pi,\pi)\), and edge probabilities
\[
 p^\star_{ij}=\sigma\{-2.8+1.5\cos(t_i-t_j)\},\qquad i<j.
\]
Node \(i\), indexed from zero, belongs to group \(i\bmod K\) for \(K=4,8\). This assignment gives two balanced partitions independent of the latent positions. The eight-group partition refines the four-group partition. Evaluating each graph under both partitions gives eighty graph--partition cases.

Off-diagonal generating probabilities range from 0.013 to 0.214 and average 0.086. The realized graphs contain 51--82 edges, with mean 67.45 and mean degree 3.37. At \(K=4\), the mean count is 0.84 and 41.7\% of counts are zero. At \(K=8\), the mean is 0.42 and 64.6\% are zero. We developed the count diagnostics, conditional references and stability checks after examining the initial results for these forty graphs. Section~\ref{sec:fresh} evaluates the resulting comparisons on an independent sample.

Both estimators use the observed \(Y,B\). The circular fit uses the generating model form and fixes the interaction strength at its true value \(\zeta=1.5\). The intercept has prior \(\alpha\sim N(-2,2^2)\), and the angles have independent uniform priors. Each of four chains runs for 1,600 warmup and 3,200 retained sweeps. Two chains start from independent uniform angles and an intercept drawn from \(N(-2,1^2)\). The third starts with equally spaced angles assigned in random order and \(\alpha=-4\). The fourth starts with \(\alpha=0\) and two angular clusters centered at 0 and \(\pi\), perturbed with standard deviation 0.1 radians and randomly assigned to nodes. Initialization uses neither the true angles nor the edges.

Each sweep proposes a normal random-walk move for the intercept, followed by wrapped-normal moves for the angles in random node order. The initial proposal scales are 0.2 for the intercept and 0.7 for the angles. Every 50 warmup sweeps, each scale is multiplied by \(\exp\{\operatorname{clip}(a_{\rm acc}-0.35,-0.3,0.3)\}\) and then clipped to \([0.02,3]\), where $a_{\rm acc}$ is the corresponding window acceptance rate. Adaptation ends after warmup. The fitted matrix averages probabilities across all 12,800 retained draws. We assess mixing for each of the 780 upper-triangular probabilities, the intercept, the working log-likelihood and the mean squared count residual. Following \citet{vehtari2021rank}, we require the larger of the rank-normalized and folded split \(\widehat R\) values to be below 1.01 and the bulk effective sample size (ESS) to be at least 400. For the MCSE of each probability mean, we use the ESS for the mean on the original probability scale.

For the nuclear fit, a consensus ADMM solver splits \(P\) into a nuclear-norm variable \(L\) and a constrained variable \(C\). With scaled dual matrices $U,V$ (distinct from the pair sets $\mathcal U,\mathcal V$) and $\rho=2$, it alternates
\begin{align}
 P&\leftarrow(YB^\top+\rho(L-U+C-V))(BB^\top+2\rho I)^{-1},\\
 L&\leftarrow\operatorname{SVT}_{\lambda/\rho}(P+U),\\
 C&\leftarrow\Pi_{\mathcal C}(P+V),
\end{align}
then updates \(U\leftarrow U+P-L\) and \(V\leftarrow V+P-C\). The singular value thresholding operator \(\operatorname{SVT}\) subtracts the indicated threshold from each singular value, truncating at zero. The projection \(\Pi_{\mathcal C}\) symmetrizes the matrix, clips entries to \([0,1]\) and sets the diagonal to zero. All variables start at zero for each fit. The primal and dual residuals are
\[
 r=\{\|P-L\|_F^2+\|P-C\|_F^2\}^{1/2},
 \qquad s=\rho\|(L-L_{\rm old})+(C-C_{\rm old})\|_F.
\]
For each of the three penalties, the solver stops when \(r,s<10^{-7}\) or after 50,000 iterations, whichever comes first, and returns the feasible matrix \(C\). An earlier checkpoint on the same trajectory records the first iterate satisfying the \(10^{-6}\) criterion, or iteration 10,000 if it comes first.

For each eight-group graph, exhaustive enumeration covers all 28 between-group and eight within-group blocks. The largest block support contains 58 configurations. The conditional references use floating-point weights on these finite supports. An independent implementation uses edge-fixed dynamic programming for between-group margins and a bitmask calculation with Bernoulli likelihoods for within-group blocks. The two implementations agree exactly on the uniform marginals, forced sets and support counts. Their oracle marginals differ by at most \(2.8\times10^{-15}\), and the maximum absolute error in \(QB=Y\) is below \(6.3\times10^{-15}\).

\subsection{Generating-probability error and raw ranking}
The nuclear fit at \(\lambda=1\) has higher mean raw AUC and lower mean count RMSE than the circular fit under both partitions (Table~\ref{tab:estimators}). For the generating probabilities, the circular fit has the lower RMSE. Its mean RMSE is 1.16\% below the density baseline at \(K=4\) and 2.41\% below at \(K=8\), with paired mean differences of \(-0.000828\) (MCSE 0.000140) and \(-0.001719\) (MCSE 0.000328). Nuclear RMSE is about 1.75 and 1.87 times the density baseline at the two partitions, respectively.

\begin{table}[htbp]
 \centering\small
 \setlength{\tabcolsep}{4pt}
 \input{tables/estimator-results.tex}
 \caption{Estimator performance under the two partitions. Entries give the mean (MCSE) across the forty exploratory sparse circular graphs. Circular estimates average draws from four chains, and nuclear estimates use the final \(\lambda=1\) iterates.}
 \label{tab:estimators}
\end{table}

Zero counts alone exclude an average of 70.3\% of nonedges at \(K=4\) and 91.5\% at \(K=8\). The count product, which assigns zero to every excluded pair, has mean raw AUC 0.908 and 0.970, respectively. At \(K=8\), applying the same zero rule to the nuclear score raises its mean AUC to approximately 0.976. On the remaining pairs in \(\mathcal U\), nuclear minus count product has a paired mean AUC difference of approximately 0.061.

\subsection{Complete constraints and conditional references}
The eight-group constraints fix an average of 24.95 present edges per graph, or 37.0\% of realized edges on average. They also fix an average of 655.65 nonedges, of which only 3.75 per graph are additional to zero exclusion. Of the 780 pairs, an average of 99.4 remain in \(\mathcal V\), comprising 42.5 edges and 56.9 nonedges.

On \(\mathcal V\), the uniform reference has mean AUC 0.680, followed by the direct nuclear score at \(\lambda=1\) (0.643), the count product (0.613) and the direct circular score (0.594). Nuclear AUC falls below uniform on 39 of forty graphs, with a paired mean difference of \(-0.038\) (MCSE 0.00402). Its mean advantage over the count product narrows to about 0.030 when the comparison excludes all forced pairs (Table~\ref{tab:ranking}).

Weighting the same feasible graphs by the generating odds from \(P^\star\) gives the conditional oracle a mean AUC of 0.889 within \(\mathcal V\). Its advantage over uniform is positive on all forty graphs and averages 0.209 (MCSE 0.01179).

Full-pair AUC gives little weight to ranking within \(\mathcal V\) in these graphs. The weight \(m_{\mathcal V}n_{\mathcal V}/(mn_0)\) in \eqref{eq:full-decomposition} averages 0.05176. A score that orders the forced states correctly and ties every pair in \(\mathcal V\) already attains mean full-pair AUC 0.974. The uniform reference and the conditional oracle attain 0.984 and 0.995.

The uniform reference also has lower mean Brier loss for reconstruction of \(A\) on \(\mathcal V\) than either direct fitted score (Table~\ref{tab:risk}). Its mean conditional excess Brier risk is 0.089, approximately 0.031 below nuclear at \(\lambda=1\).

\begin{table}[htbp]
 \centering\small
 \input{tables/conditional-risk.tex}
 \caption{Graph-prediction loss within \(\mathcal V\) at \(K=8\), mean (MCSE) across the forty exploratory sparse circular graphs. Realized Brier loss uses \(A\) as its target. Conditional excess Brier risk measures mean squared distance to \(Q^{\mathrm{or}}\) as in \eqref{eq:brier}. Nuclear uses \(\lambda=1\).}
 \label{tab:risk}
\end{table}

\subsection{Penalty and computation sensitivity}\label{app:exploratory-checks}
Across the three prescribed nuclear penalties, stronger regularization lowers both mean generating-probability RMSE and mean AUC (Table~\ref{tab:penalty}). All three penalties give higher mean RMSE than the density baseline. At \(K=8\), the paired mean AUC differences on \(\mathcal V\) for direct nuclear scoring relative to uniform are \(-0.016\) (MCSE 0.00458), \(-0.038\) (0.00402) and \(-0.102\) (0.00568) for \(\lambda=0.25,1,4\), respectively.

\begin{table}[htbp]
 \centering\footnotesize
 \setlength{\tabcolsep}{3pt}
 \input{tables/penalty.tex}
 \caption{Nuclear penalty sensitivity, mean (MCSE) across the forty exploratory sparse circular graphs, including all final iterates. The last column counts fits that reached 50,000 iterations before satisfying both \(10^{-7}\) residual tolerances. Full-constraint enumeration and \(\operatorname{AUC}_{\mathcal V}\) are available only at \(K=8\).}
 \label{tab:penalty}
\end{table}

The sign of the mean raw AUC difference between nuclear and circular depends on the penalty. At \(\lambda=1\), nuclear minus circular averages 0.045 (MCSE 0.00424) for \(K=4\) and 0.066 (MCSE 0.00367) for \(K=8\). At \(\lambda=4\), the corresponding differences are \(-0.005\) (0.00469) and \(-0.008\) (0.00441), both within two MCSEs of zero. Moving from four to eight groups also increases nuclear probability RMSE at the two smaller penalties and decreases it at the largest.

At \(\lambda=1\), the earlier checkpoints reproduce the saved estimates and solver diagnostics exactly, including six fits that reached 10,000 iterations under the \(10^{-6}\) residual rule. Continuing to the stricter rule changes mean raw AUC by \(3.42\times10^{-6}\) and \(-8.67\times10^{-7}\) at \(K=4,8\), respectively. The changes in mean generating-probability RMSE are smaller.

Five of the 240 final fits reached the 50,000-iteration cap. Four occurred at \(K=4\), for seed 1009 at \(\lambda=4\), seed 1010 at \(\lambda=1\), and seed 1019 at both \(\lambda=1\) and \(\lambda=4\). The fifth occurred at \(K=8\) for seed 1006 at \(\lambda=4\). Their primal residuals range from \(1.25\times10^{-7}\) to \(2.44\times10^{-6}\), although all five dual residuals are below \(10^{-7}\). The reported comparisons include the final feasible matrices from these five capped fits.

All eighty four-chain circular fits meet the matrix and scalar diagnostic thresholds. Across the 62,400 probability entries, the maximum \(\widehat R\) is 1.00398, the minimum bulk ESS is 2321 and the largest MCSE for a probability mean is 0.00160.

We compared graph-paired performance contrasts across chain configurations and subsets of the retained draws. The two criteria were an absolute change exceeding half the larger of the two graph-level MCSEs and a reversal of sign for a contrast previously more than two MCSEs from zero. Neither occurred when comparing the two halves of the retained draws or chains 1--2 with chains 3--4. The circular fit's small RMSE advantage over density held in both comparisons. The earlier runs used two chains with 400 warmup and 800 retained sweeps per chain. With four longer chains and the additional starts, mean raw AUC rose from 0.837 to 0.847 at \(K=4\) and from 0.897 to 0.901 at \(K=8\). Changes in ranking contrasts exceeded the first criterion. We report the four-chain estimates throughout.

\subsection{Reproducibility}
The \texttt{exploratory/} directory of the \hyperref[sec:data]{replication archive} contains the exploratory inputs, conditional probabilities and support counts, early and final iterates from each nuclear fit, circular mean matrices and scalar traces, diagnostic arrays, and executable source. It includes separate checks of conditional marginals, integer support counts, saved matrix means, solver objectives, ranking metrics and paired summaries. Its code can regenerate the approximately 6.4 GB of exploratory probability draws for the draw-level checks. The original inputs and earlier computation are preserved in the \href{https://github.com/kwt-klure/ard-reconstruction-reproducibility/tree/9fa2b75305216b279cb03501c0d7a56b653a3ae8}{GitHub repository \nolinkurl{kwt-klure/ard-reconstruction-reproducibility}} at commit \texttt{9fa2b7530521}.

The reported AUCs give half credit for exact numerical ties. We also checked uniform-reference ties by integer cross multiplication. For the main \(K=8\) comparison, using the four-chain circular fit and the final \(\lambda=1\) nuclear fit, tie tolerances of \(10^{-12}\) and \(10^{-10}\) changed mean AUC by less than \(10^{-5}\). For each conditional reference, AUC on \(\mathcal V\) was identical under all three tie rules. The archived outputs include floating-point and runtime details. The comparisons include every prescribed graph, fit and chain.

%% file: tables/estimator-results.tex
\begin{tabular}{llrrr}
\toprule
$K$ & Estimator & $P^\star$ RMSE & Raw AUC & Count RMSE \\
\midrule
4 & Circular & 0.0704 (0.00019) & 0.847 (0.0043) & 0.8241 (0.00902) \\
4 & Nuclear ($\lambda=1$) & 0.1247 (0.00115) & 0.891 (0.0034) & 0.1964 (0.00175) \\
4 & Density & 0.0712 (0.00020) & 0.500 (0.0000) & 0.8756 (0.01069) \\
\midrule
8 & Circular & 0.0695 (0.00029) & 0.901 (0.0039) & 0.5801 (0.00485) \\
8 & Nuclear ($\lambda=1$) & 0.1330 (0.00125) & 0.967 (0.0014) & 0.2107 (0.00152) \\
8 & Density & 0.0712 (0.00020) & 0.500 (0.0000) & 0.6220 (0.00616) \\
\bottomrule
\end{tabular}

%% file: tables/conditional-risk.tex
\begin{tabular}{lrr}
\toprule
Probability matrix & Realized Brier on $\mathcal{V}$ & Conditional excess Brier on $\mathcal{V}$ \\
\midrule
$Q^{\mathrm{unif}}$ & 0.2182 (0.00141) & 0.0887 (0.00438) \\
$Q^{\mathrm{or}}$ & 0.1300 (0.00655) & 0.0000 (0.00000) \\
True $P^\star$ & 0.3250 (0.00346) & 0.1954 (0.00495) \\
Circular & 0.3505 (0.00424) & 0.2208 (0.00635) \\
Nuclear ($\lambda=1$) & 0.2496 (0.00239) & 0.1199 (0.00501) \\
\bottomrule
\end{tabular}

%% file: tables/penalty.tex
\begin{tabular}{rrrrrrr}
\toprule
$K$ & $\lambda$ & $P^\star$ RMSE & Raw AUC & AUC on $\mathcal{U}$ & AUC on $\mathcal{V}$ & Final caps \\
\midrule
4 & 0.25 & 0.1436 (0.00135) & 0.907 (0.0033) & 0.710 (0.0041) & -- & 0 \\
4 & 1 & 0.1247 (0.00115) & 0.891 (0.0034) & 0.693 (0.0042) & -- & 2 \\
4 & 4 & 0.1007 (0.00087) & 0.842 (0.0040) & 0.653 (0.0046) & -- & 2 \\
\midrule
8 & 0.25 & 0.1723 (0.00161) & 0.980 (0.0011) & 0.773 (0.0071) & 0.665 (0.0061) & 0 \\
8 & 1 & 0.1330 (0.00125) & 0.967 (0.0014) & 0.715 (0.0074) & 0.643 (0.0066) & 0 \\
8 & 4 & 0.0915 (0.00062) & 0.894 (0.0024) & 0.582 (0.0076) & 0.578 (0.0066) & 1 \\
\bottomrule
\end{tabular}

%% file: numerical-details.tex
\section{Numerical details for the fresh studies and plug-in comparison}\label{app:independent-checks}
\subsection{Fresh circular study}\label{app:density-checks}
The sparse, 15\% and 25\% settings use intercepts of $-2.8$, $-2.103309259462$ and $-1.369700517844$. The interaction strength is $\zeta=1.5$ at all three densities.

The circular prior, working likelihood, starting configurations, proposals and warmup adaptation follow Appendix~\ref{app:exploratory-protocol}. Each fit averages 12,800 probability matrices from four chains with 1,600 warmup and 3,200 retained sweeps per chain. The nuclear penalties are $\lambda\in\{0.25,1,4\}$, with $\lambda=1$ the main reference.

Before generating the fresh sample, we fixed an ADMM solver with the exploratory objective, initialization and residual tolerance and with finite residual balancing as in \citet[sec.~3.4.1]{boyd2011admm}. The solver starts at $\rho=2$. Every 25 iterations it doubles $\rho$ if $r>10s$ and halves it if $s>10r$, keeping $\rho$ within $[0.125,128]$. After each change it multiplies the scaled dual variables by the ratio of the old $\rho$ to the new $\rho$. Adaptation ends at iteration 100,000. The solver stops when both residuals are below $10^{-7}$ or after 200,000 iterations, whichever comes first. Applied to the 240 exploratory graph--partition--penalty cases, this procedure met both tolerances in every case and changed mean raw AUC by less than $2.5\times10^{-6}$ in absolute value in every partition--penalty setting.

All 240 circular fits meet the matrix and scalar diagnostic thresholds. Of 720 nuclear fits, 719 satisfy both residual tolerances. The sparse-density (8.53\%) fit for seed 21021 at $K=4$, $\lambda=4$ reached the 200,000-iteration cap. Its primal residual was $2.26827625\times10^{-7}$ and its dual residual was $4.91301071\times10^{-8}$. The comparisons include its final feasible matrix along with all other prescribed fits and chains.

For all 120 graphs, independent fixed-margin calculations reproduce the uniform marginals exactly and the oracle marginals to within $4.9961\times10^{-15}$. Further checks cover the saved objectives, all retained probability-matrix means, scalar draws and selected direct calculations of $\widehat R$. For ESS and ADMM trajectories, the record uses the original computations. The fresh-study record links the tables to graph-level values and source hashes, including the code for the tie-tolerance checks. The \texttt{studies/fresh/} directory of the \hyperref[sec:data]{replication archive} contains the inputs and fit summaries. Its reproduction guide explains how to regenerate the approximately 19.2 GB of probability draws from all 960 chains.

\subsection{Stochastic blockmodel study}\label{app:sbm-checks}
The blockmodel matches the dyad mean and variance of the circular edge probabilities at the 15\% intercept in Appendix~\ref{app:density-checks} and $\zeta=1.5$. With $\mu_P=0.15$, $\omega=3/13$ and $v=0.013569347207664$, equation~\eqref{eq:sbm-match} gives $p_{\mathrm{in}}=0.362676179575618$ and $p_{\mathrm{out}}=0.086197146127314$.

All 120 nuclear fits meet both residual tolerances, and the four-chain circular fit meets the diagnostic thresholds on all forty graphs. Independent calculations reproduce the uniform marginals exactly on every graph. The largest discrepancy for an oracle marginal is $4.774\times10^{-15}$. Tie tolerances of $10^{-12}$ and $10^{-10}$ change mean AUC by at most $4.95\times10^{-5}$ and preserve the direction of the principal comparisons. The checks include saved probability means and selected direct calculations of $\widehat R$. As in the fresh circular study, ESS and ADMM trajectories come from the original computations. The \texttt{studies/sbm/} directory of the \hyperref[sec:data]{replication archive} contains the simulation parameters, inputs and fitted matrices.

\subsection{Plug-in conditional comparison}\label{app:plugin-checks}
For all 200 eight-group cases, we enumerated the existing block supports and reweighted them directly from the saved fits, including the capped exploratory nuclear fit at $\lambda=4$. Each graph yields one circular plug-in matrix and nine nuclear plug-in matrices, one for each combination of three fixed penalties and three fixed clipping thresholds. All 2,000 matrices preserve the original forced states, and the largest absolute count residual is $1.33\times10^{-15}$. Conditioning on the same counts with true $P^\star$ as input reproduces all 200 saved oracle matrices to within $2.34\times10^{-15}$ entrywise. Separate whole-graph Bernoulli-product calculations on small finite supports check the weighted law, additive cancellation and zero-probability boundary cases.

At the primary threshold $\epsilon=10^{-6}$, clipping alone leaves every reported mean nuclear AUC on $\mathcal V$ unchanged. Across $\epsilon=10^{-4},10^{-6},10^{-8}$, the largest changes from the primary threshold in a study setting are $1.56\times10^{-5}$ in mean plug-in AUC and $1.03\times10^{-6}$ in mean conditional excess Brier risk. Individual conditional probabilities change by as much as $6.71\times10^{-4}$. The archived source and outputs include graph-level clipping counts, raw scores and clipping-only scores.

The reported AUCs use the stored conditional probabilities and give half credit for exact numerical ties. An earlier calculation used log marginal odds, whose rounding could separate tied interior probabilities or introduce ties between distinct probabilities. The two calculations differ in mean AUC by at most $8.42\times10^{-6}$ in any setting and agree on every circular result. The archived log marginal odds serve as a numerical diagnostic. There was no underflow in the weighted sums, and every unresolved marginal remained strictly between zero and one after rounding. Probability-tie tolerances of $10^{-12}$ and $10^{-10}$ change mean plug-in AUC by at most $2.64\times10^{-4}$ in any setting and preserve the direction of the reported mean comparisons. Independent checks of the saved matrices reproduce all scores, graph-paired contrasts and MCSEs, using direct pairwise comparisons for AUC. Source hashes agree before and after the calculation.